\documentclass[aps,prl,twocolumn,superscriptaddress,longbibliography]{revtex4-2}
\usepackage[utf8]{inputenc}
\usepackage[T1]{fontenc}

\usepackage[english]{babel}
\usepackage{graphicx}  
\usepackage{dcolumn}   
\usepackage{bm}        
\usepackage{amssymb}   
\usepackage{amsmath, latexsym}
\usepackage{units}
\usepackage[dvipsnames]{xcolor}
\usepackage[%
colorlinks=true,
urlcolor=RoyalBlue,
linkcolor=RoyalBlue,
citecolor=RoyalBlue,
]{hyperref}
\usepackage{etoolbox}
\usepackage{xfrac}
\usepackage{orcidlink}

\usepackage{sidecap}
\sidecaptionvpos{figure}{t}

\usepackage[type1]{libertine}                                        
\usepackage{textcomp}
\usepackage[scaled=.85]{beramono}
\usepackage[libertine,cmintegrals,cmbraces,vvarbb,slantedGreek]{newtxmath}
\usepackage[scr=boondoxo]{mathalfa}
\usepackage{bm}
\usepackage[lf]{carlito}

\makeatletter
\DeclareRobustCommand{\cev}[1]{%
  \mathpalette\do@cev{#1}%
}
\newcommand{\do@cev}[2]{%
  \fix@cev{#1}{+}%
  \reflectbox{$\m@th#1\vec{\reflectbox{$\fix@cev{#1}{-}\m@th#1#2\fix@cev{#1}{+}$}}$}%
  \fix@cev{#1}{-}%
}
\newcommand{\fix@cev}[2]{%
  \ifx#1\displaystyle
    \mkern#23mu
  \else
    \ifx#1\textstyle
      \mkern#23mu
    \else
      \ifx#1\scriptstyle
        \mkern#22mu
      \else
        \mkern#22mu
      \fi
    \fi
  \fi
}
\makeatother

\usepackage{vmargin}
\setpapersize{A4} \oddsidemargin20mm \textwidth170mm \topmargin10mm
\newcommand{\figref}[2][]{\def\a{#1}\def\e{}Fig.\ \ref{#2}\if\a\e\else (\a)\fi}
\newcommand{\Figref}[2][]{\def\a{#1}\def\e{}Figure \ref{#2}\if\a\e\else (\a)\fi}

\newcommand{\panelcaption}[1]{(#1)}
\newcommand{\panelsubcaption}[1]{(#1)}

\begin{document}

\title{Continuous Tuning of the Charge-Phase Uncertainty in a Josephson Junction}

\author{Irena Padniuk}
\affiliation{Max-Planck-Institut f\"ur Festk\"orperforschung, Heisenbergstraße 1, 70569 Stuttgart, Germany}
\author{Xianzhe Zeng}
\affiliation{Max-Planck-Institut f\"ur Festk\"orperforschung, Heisenbergstraße 1, 70569 Stuttgart, Germany}
\author{Juan Carlos Cuevas\,\orcidlink{0000-0001-7421-0682}}
\affiliation{Departamento de F\'{\i}sica Te\'{o}rica de la Materia Condensada,
Universidad Aut\'{o}noma de Madrid, E-28049 Madrid, Spain}
\affiliation{Condensed Matter Physics Center (IFIMAC), Universidad Aut\'{o}noma de Madrid, E-28049 Madrid, Spain}
\author{Joachim Ankerhold\,\orcidlink{0000-0002-6510-659X}}
\affiliation{Institute for Complex Quantum Systems and IQST, Universität Ulm, Albert-Einstein-Allee 11, 89069 Ulm, Germany}
\author{Klaus Kern\,\orcidlink{0000-0002-1785-7874}}
\affiliation{Max-Planck-Institut f\"ur Festk\"orperforschung, Heisenbergstraße 1, 70569 Stuttgart, Germany}
\affiliation{Institut de Physique, Ecole Polytechnique F{\'e}d{\'e}rale de Lausanne, 1015 Lausanne, Switzerland}
\author{Christian R. Ast\,\orcidlink{0000-0002-7469-1188}}
\email[Corresponding author; electronic address:\ ]{c.ast@fkf.mpg.de}
\affiliation{Max-Planck-Institut f\"ur Festk\"orperforschung, Heisenbergstraße 1, 70569 Stuttgart, Germany}

\date{\today}

\begin{abstract}
Quantum mechanics constrains conjugate observables from being simultaneously measurable with arbitrary precision. In a Josephson junction, these are the transferred electric charge and the quantum-mechanical phase difference between the superconducting domains. Which of them fluctuates determines the supercurrent: a dissipative trickle of single Cooper pairs in one limit, a coherent dissipationless flow in the other. Bridging both regimes in one device remained elusive because the Josephson and charging energies are fixed at fabrication. Here, we use the tunable tunnel junction of a scanning tunneling microscope at millikelvin temperature to vary their ratio continuously over many orders of magnitude. In this way, we monitor the smooth transition between incoherent and coherent Cooper pair flow in a single junction, revealing the quantum-to-classical transition in a controlled way.
\end{abstract}

\maketitle

In quantum mechanics, conjugate observables such as position and momentum are bound together by Heisenberg's uncertainty principle: they cannot be measured simultaneously with arbitrary precision, a complementarity that has no counterpart in classical physics. For superconducting tunnel junctions, the basic elements of superconducting quantum computing platforms~\cite{nakamura_coherent_1999,vion_manipulating_2002,devoret_superconducting_2013,kjaergaard_superconducting_2020,blais_circuit_2021}, the charge transferred across the junction takes the role of momentum and the superconducting phase difference that of position. In an actual circuit the two cannot fluctuate independently: if one follows a well-defined motion, the other varies strongly. Which of them dominates is controlled by the ratio of the two corresponding energy scales, the Josephson energy $E_\mathrm{J}$ and the charging energy $E_\mathrm{C}$. If $E_\mathrm{J}$ dominates, phase fluctuations are strongly suppressed and the junction sustains a coherent, dissipationless flow of Cooper pairs (DCT), as first predicted by Josephson~\cite{josephson_possible_1962}. In the opposite limit, phase fluctuations are strong, the charge becomes the relevant degree of freedom, and Cooper pairs cross the junction only incoherently through the exchange of energy quanta with the electromagnetic environment, a phenomenon known as dynamical Coulomb blockade (DCB)~\cite{devoret_effect_1990,averin_incoherent_1990,ingold_charge_1992,ingold_cooperpair_1994}.

In superconducting circuits designed for technological applications, the ratio $E_\mathrm{J}/E_\mathrm{C}$ is routinely fixed by design, for example to reduce the sensitivity of transmon qubits to environmental electrical noise~\cite{wallraff_strong_2004,koch_chargeinsensitive_2007,bouchiat_quantum_1998}. Therefore, comparing the two domains usually means comparing different devices with uncontrolled stray capacitances and environmental impedances. How a junction evolves, when this ratio is instead tuned continuously \textit{in situ} from one limiting domain to the other, remains open, both experimentally and theoretically, as does the related evolution from the few-channel DCB regime to the many-channel DCT regime. This is of direct relevance to superconducting quantum devices and parametric amplifiers, in particular when some transport channels are highly transparent~\cite{willsch_observation_2024}.

Theoretically, the coherent regime is well described by the resistively shunted junction (RSJ) model and its extensions~\cite{stewart_currentvoltage_1968,mccumber_effect_1968} and the Coulomb blockade regime by $P(E)$-theory~\cite{devoret_effect_1990,averin_incoherent_1990,ingold_charge_1992,ingold_cooperpair_1994}, but a unified framework remains elusive~\cite{ingold_cooperpair_1994,schon_quantum_1990}. Some progress has been made for a Josephson junction with many transmission channels in an ohmic environment, for which a generalization of the RSJ model has been formulated~\cite{grabert_phase_1998}. The environment is expected to lose its influence towards the coherent supercurrent domain~\cite{caldeira_quantum_1983}, but general results are missing because perturbative treatments in either $E_\mathrm{J}$ or $E_\mathrm{C}$ fail.

Here we overcome the experimental limitations using the tunnel junction of a scanning tunneling microscope (STM)~\cite{naaman_fluctuation_2001,jack_critical_2016,ast_sensing_2016,randeria_scanning_2016,roychowdhury_microwave_2015}, in which the tip-sample distance provides a continuously tunable knob for the tunnel coupling and hence $E_\mathrm{J}$, while $E_\mathrm{C}$ and the electromagnetic environment remain essentially unchanged, as shown in the inset of \figref[a]{Fig1} along with the full data set. We thereby sweep $E_\mathrm{J}/E_\mathrm{C}$ over several orders of magnitude in a single junction, tracing the full evolution of the current--voltage characteristics from incoherent Cooper pair tunneling in the DCB regime to a coherent Josephson supercurrent in the DCT regime.

\begin{figure*}
\centering
\includegraphics[width=0.9\textwidth]{figure1_arxiv.png}
\caption{\label{Fig1} \panelcaption{a} Tunneling current vs.\ voltage for different values of the junction conductance $0.007\,G_0\leq G_\text{N} \leq 28.5\,G_0$, normalized to $G_\text{N}$. The coherence peaks are visible at $\pm1.46\,$mV. The offset in the normal-state regime is due to the excess current; the steps inside the gap are from multiple Andreev reflections. The Josephson effect at high conductance is visible near zero voltage. The lower inset shows the tunneling current normalized to $G_\text{N}^2$ to highlight the low-conductance Josephson effect. The upper inset shows a schematic of the tunnel junction in the DCB and the DCT regime. \panelcaption{b} Schematic evolution of the Josephson effect as function of the ratio $E_\mathrm{J}/E_\mathrm{C}$. At low conductance in the DCB regime, the maximal Josephson current (switching current) evolves quadratically with $G_\text{N}$; at high conductance in the DCT regime, it evolves linearly. Since the critical current cannot be exceeded, the evolution must transition from quadratic to linear at sufficiently large $E_\mathrm{J}/E_\mathrm{C}$. \panelcaption{c} Evolution of the Josephson current within a voltage window of $\pm 20\,\upmu$V as function of $G_\text{N}$ for selected $I(V)$-curves from \panelsubcaption{a}. The change in the slope at zero voltage and the reduced interaction with the environment are clearly visible.}
\end{figure*}

Both limits are individually well understood. In the DCB regime, $P(E)$-theory predicts a Josephson current that scales with the square of the normal-state conductance, $I_\text{J}^\text{DCB}\propto G_\text{N}^2$, whereas in the DCT regime it is linear, $I_\text{J}^\text{DCT}\propto G_\text{N}$, set by the critical current $I_\text{C} = \frac{2e}{\hbar}E_\mathrm{J}$. This critical current cannot be exceeded and, for a conventional junction with a sinusoidal current-phase relation, grows linearly with $G_\text{N}$~\cite{josephson_possible_1962,ambegaokar_tunneling_1963}. The same linear dependence holds for the switching current $I_\text{S}$, the maximal current the junction sustains in an actual circuit. The most direct way to observe the transition is therefore to follow how the quadratic dependence of $I_\text{S}$ crosses over into a linear one as $G_\text{N}$ increases when the tip approaches the sample. Since the quadratically evolving DCB current must eventually meet the linearly evolving critical current, a smooth crossover from dissipative to dissipationless Cooper pair tunneling is expected, as sketched in \figref[b]{Fig1}.

\sidecaptionvpos{figure}{c}
\begin{SCfigure*}
\centering\includegraphics[width=1.4\columnwidth]{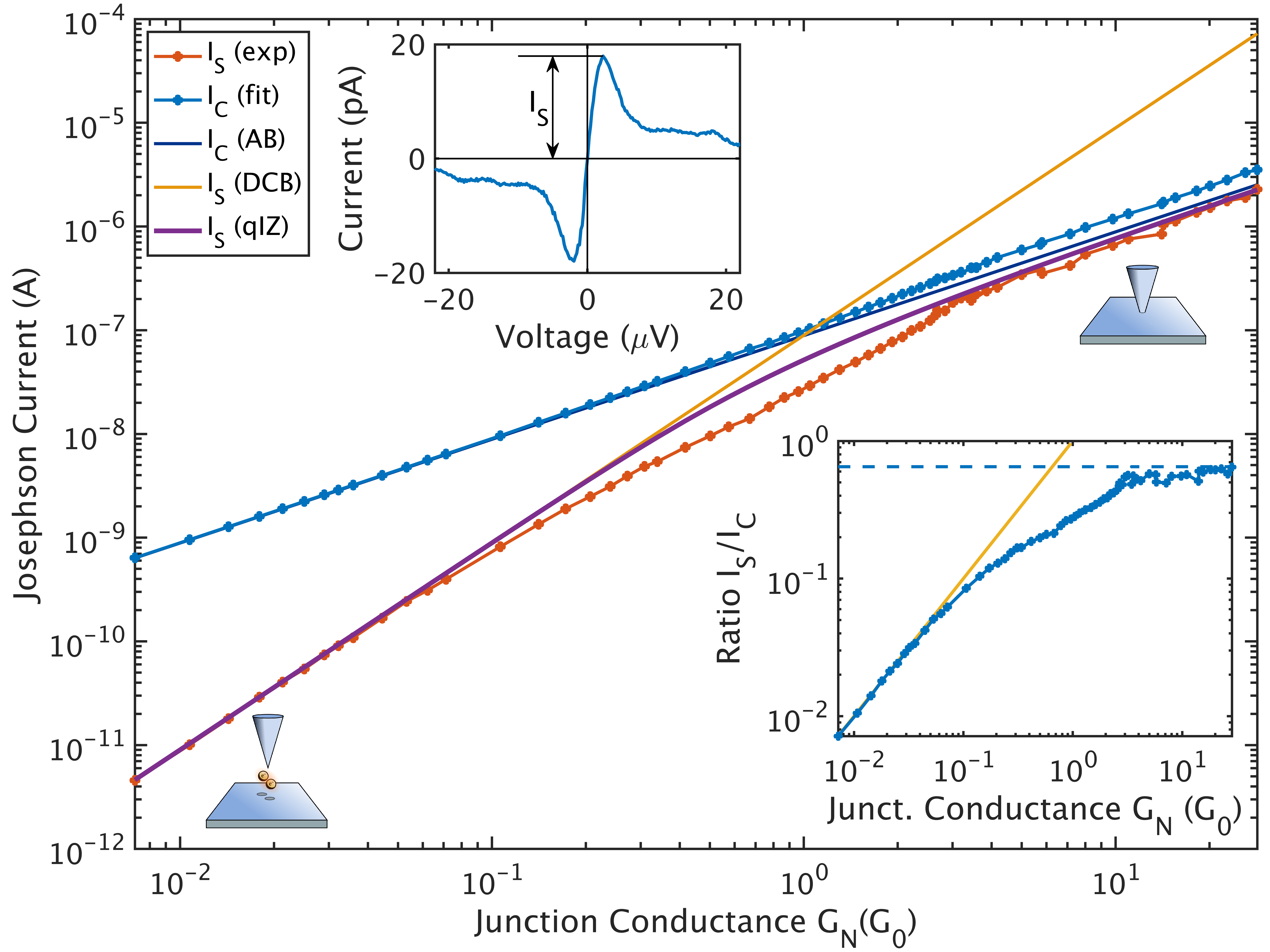}
\caption{\label{Fig2} Josephson current as function of junction conductance. The critical current $I_\text{C}$ evolves linearly and is calculated both from the Ambegaokar-Baratoff (AB) formula (blue line) and from the channel-resolved fit to the experimental data (blue line with markers). The switching current $I_\text{S}$ initially evolves quadratically and then linearly, both for the experimental data (red line with markers) and for the qIZ model (purple line). $I_\text{S}$ has been extracted from the $I(V)$-curves as illustrated in the upper inset. The yellow line shows the quadratic evolution expected from the DCB model. The experimental $I_\text{S}$ is in good agreement with the qIZ model in the extreme limits; in the transition regime, deviations are visible that may be traced to the simplifying assumptions of the model. The lower inset shows the experimentally extracted ratio $I_\text{S}/I_\text{C}$. The ratio initially evolves linearly (cf.\ yellow line) and saturates at about $67\%$.}
\end{SCfigure*}

Realizing this crossing within an experimentally accessible conductance range is the central challenge; it is eased by lowering the temperature and improving the energy resolution, which shifts the crossover to lower conductance. We use a unique STM operating at a base temperature of 10\,mK with energy resolution optimized through internal filtering~\cite{assig_10_2013,zeng_enhancing_2026}, a vanadium tip, and a V(100) sample, both superconducting with gap parameters $\Delta = 700\,\upmu$eV (tip) and $750\,\upmu$eV (sample) (see SM). The Josephson current was measured for junction conductances between $0.007\,G_0$ and $28.5\,G_0$, where $G_0 = 2e^2/h$ is the conductance quantum. A close-up of selected $I(V)$-curves within $\pm20\,\upmu$V is shown in \figref[c]{Fig1}, each normalized to its maximum (analysis in the SM). Their shape changes systematically with conductance. In the low-conductance DCB regime, we see the well-known $P(E)$ peak structure with a steep but finite zero-bias slope (zero-bias conductance $G_\text{ZB}$) and resonances at higher voltages from the environmental impedance. Around $G_0$, a transition regime sets in where $G_\text{ZB}$ increases continuously and the environmental resonances disappear. At the highest conductances, $G_\text{N} > 10\,G_0$, where tip and sample are in contact, $G_\text{ZB}\rightarrow \infty$ within our resolution, as expected for dissipationless Cooper pair tunneling, and the current is nearly constant at finite voltage.

In \figref{Fig2}, the switching current $I_\text{S}$ extracted from each spectrum as the peak of the supercurrent~\cite{steinbach_direct_2001} (upper inset) is plotted against $G_\text{N}$. For comparison we show the quadratic evolution expected in the DCB regime, the linear critical current $I_\text{C}$ from the Ambegaokar-Baratoff (AB) formula~\cite{ambegaokar_tunneling_1963} (assuming a sinusoidal current-phase relation), the channel-resolved $I_\text{C}$, and the qIZ prediction, all discussed below. The crossover, expected near $G_0$, indeed appears as a smooth transition from a quadratic to a linear evolution. In the DCB regime, $I_\text{S}$ is more than two orders of magnitude below $I_\text{C}$; the ratio $I_\text{S}/I_\text{C}$ rises to a saturating value of about $0.67$ in the linear regime (lower inset). The same qualitative behavior is observed in a second dataset (SM).

The evolution of $I_\text{S}$ in \figref{Fig2} is a direct, model-independent observation of the transition from a quadratic to a linear dependence on $G_\text{N}$. Comparing the limits with the established models of either regime, we conclude that the data trace a continuous transition from dissipative to dissipationless Cooper pair tunneling. The role of the environment changes fundamentally across this transition: it is essential in the DCB regime, where tunneling proceeds through energy exchange with environmental modes, whereas the current becomes dissipationless in the DCT regime. In the normal state, the DCB and its energy exchange with the environment are suppressed by a Fano factor as the transport channels become more transparent~\cite{altimiras_experimental_2007,senkpiel_dynamical_2020}, but for the Josephson junction a renormalization of the environmental impedance seems more likely~\cite{joyez_self-consistent_2013}.

\begin{SCfigure*}
\centering\includegraphics[width=1.4\columnwidth]{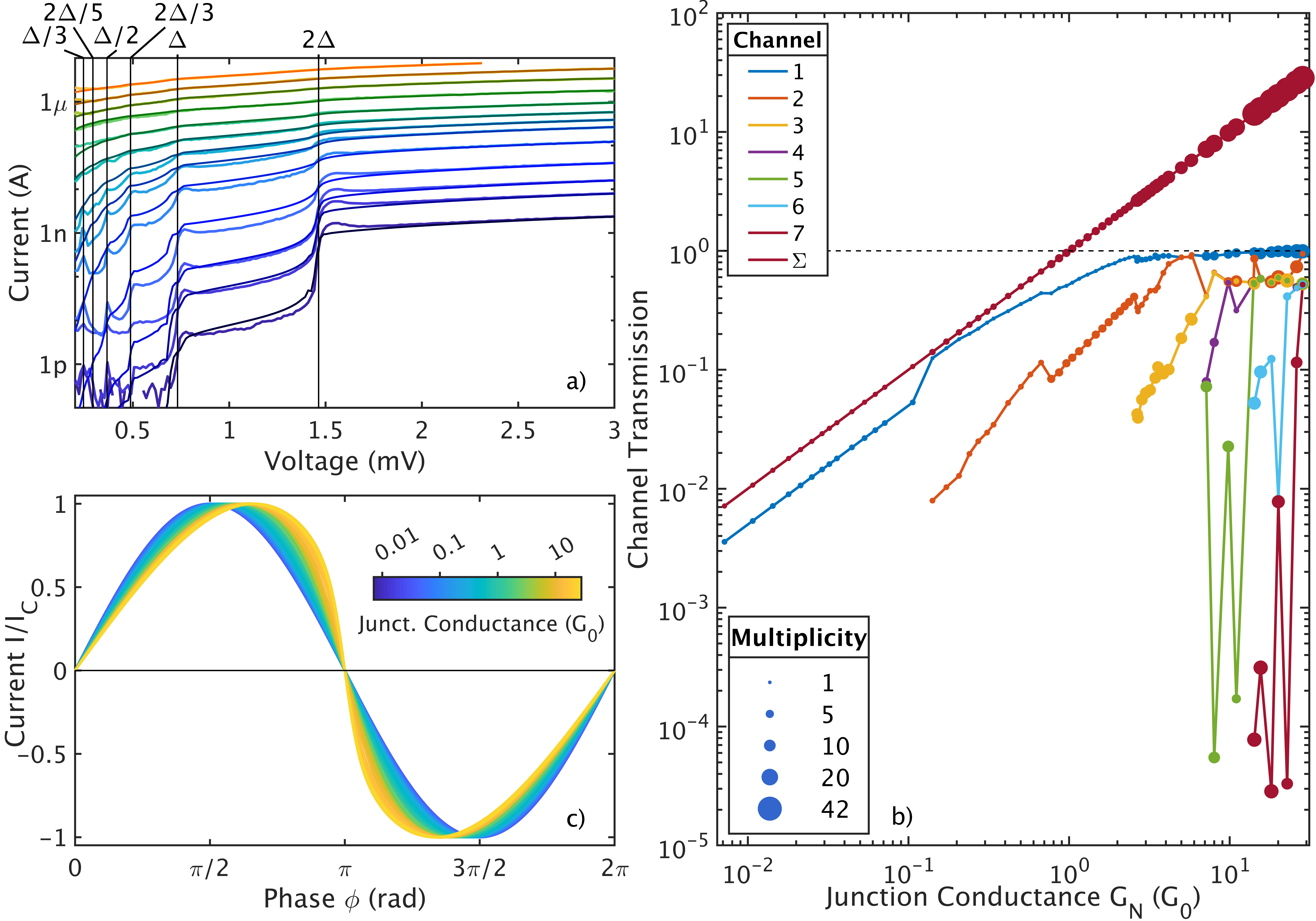}
\caption{\label{Fig3} \panelcaption{a} Fits of the quasiparticle $I(V)$-curves to a multichannel MAR model for selected values of $G_\text{N}$. The MAR onsets are indicated by vertical lines. The fits are used to extract the number of channels and their transmissions. \panelcaption{b} Channel transmissions and their multiplicity (i.e.\ how many channels share the same transmission) as function of $G_\text{N}$. The number of channels and the transmissions both grow with $G_\text{N}$, with several high-transparency channels appearing at high $G_\text{N}$. \panelcaption{c} Current-phase relation reconstructed from the channel configuration for the different values of $G_\text{N}$. The color bar encodes $G_\text{N}$ on a logarithmic scale. The current-phase relation is sinusoidal at low $G_\text{N}$ and gradually deviates from a sinusoid at high $G_\text{N}$.}
\end{SCfigure*}

To understand the junction in more detail, we go beyond the conventional sinusoidal current-phase relation and employ the more general framework of multiple Andreev reflections (MARs). Here the key properties are the number of conduction channels $n$ and their transmissions $\tau_i$, since the junction evolves from very few channels in the tunneling regime to tens of channels far beyond point contact. The channel configuration determines the current-phase relation and thus the Cooper pair tunneling. We fitted each spectrum to a multichannel MAR model (details in the SM), which reliably extracts the number of transport channels and their transmissions~\cite{cuevas_hamiltonian_1996,scheer_conduction_1997,scheer_signature_1998}. Selected $I(V)$-curves with their fits (darker color) are shown in \figref[a]{Fig3} on a logarithmic current axis, with the subgap MAR steps indicated on the top axis.

The resulting channel configuration (\figref[b]{Fig3}) gives a junction conductance $G_\text{N} = G_0 \sum_i m_i \tau_i$, where $m_i$ is the multiplicity of each channel. As the conductance increases beyond $G_0$, more channels appear and their multiplicity grows, with several reaching very high transmission ($\approx 0.96\pm 0.03$)~\cite{della_rocca_measurement_2007} near the limit $\tau = 1$ (horizontal dashed line). At the highest conductances, the junction has a total of 42 transport channels.

From the channel configuration we reconstruct the current-phase relation at each setpoint (\figref[c]{Fig3}; details in the SM), which reveals how much the junction deviates from the sinusoidal form assumed by most models, including the only theory that treats arbitrary ratios of $E_\mathrm{J}/E_\mathrm{C}$~\cite{grabert_phase_1998}. The relation is clearly sinusoidal at low conductance and gradually deviates at higher conductance. Such higher-harmonic content also appears in planar tunnel junctions (e.g.\ transmon qubits and parametric amplifiers), where it has been observed in AlO$_x$ barriers~\cite{willsch_observation_2024}. From these relations we extract the critical current $I_\text{C}$ for each setpoint (blue line with markers in \figref{Fig2}). Starting around $G_0$, it deviates from the linear AB formula~\cite{ambegaokar_tunneling_1963} and eventually saturates about 40\% higher, which we attribute to the accumulation of high-transmission channels, while still following the AB model in parallel. With several high-transmission channels present at the highest conductances, the deviation from a sinusoidal shape cannot be neglected~\cite{senkpiel_single_2020}, although the evolution remains smooth and shows no discontinuities -- a point to keep in mind when comparing experiment and theory.

As noted above, the only available approach covering arbitrary ratios of $E_\mathrm{J}/E_\mathrm{C}$ makes simplifying assumptions that are not all satisfied here, notably a sinusoidal current-phase relation and an ohmic environmental impedance~\cite{grabert_phase_1998}. We assume that these deviations are still sufficiently weak, as confirmed below (model and application details in the SM). We refer to it as the quantum Ivanchenko-Zil'berman (qIZ) model~\cite{ivanchenko_josephson_1969,grabert_phase_1998}, describing the motion of an overdamped particle in the presence of quantum-mechanical momentum fluctuations~\cite{ankerhold_strong_2001,ingold_effect_1999}. We computed a spectrum for each $G_\text{N}$, a selection of which is shown in \figref[a]{Fig4} in analogy to \figref[c]{Fig1}. The agreement with the overall shape of the $I(V)$-curves and with the measured evolution of $G_\text{ZB}$ is excellent. The higher-voltage environmental resonances are not reproduced because the impedance is reduced to an ohmic resistance $R_\text{env}$. The only parameter that varies with conductance is the Josephson energy ($E_\text{J}\propto G_\text{N}$); all others are held fixed at the values used previously~\cite{zeng_enhancing_2026} and listed in the SM, providing a consistent modeling approach.

The switching current extracted from the calculated $I(V)$-curves (purple line in \figref{Fig2}) agrees well with experiment in both extreme regimes, even though at high conductance the current-phase relation is non-sinusoidal and the critical current exceeds the AB value~\cite{ambegaokar_tunneling_1963}. Including higher harmonics in the current-phase relation and in the Josephson potential could shed more light here, but is challenging even numerically. The largest deviation is in the transition region around $G_0$, where the experimental switching current lies well below the theoretical value. This may stem from the simplified ohmic treatment of the environment~\cite{joyez_self-consistent_2013} or from the experimental circuitry, where an instability due to a negative differential conductance modifies the shape of the spectrum~\cite{ternes_scanning_2006,escribano_feedback_2025}.

\begin{figure*}
\centering\includegraphics[width=\textwidth]{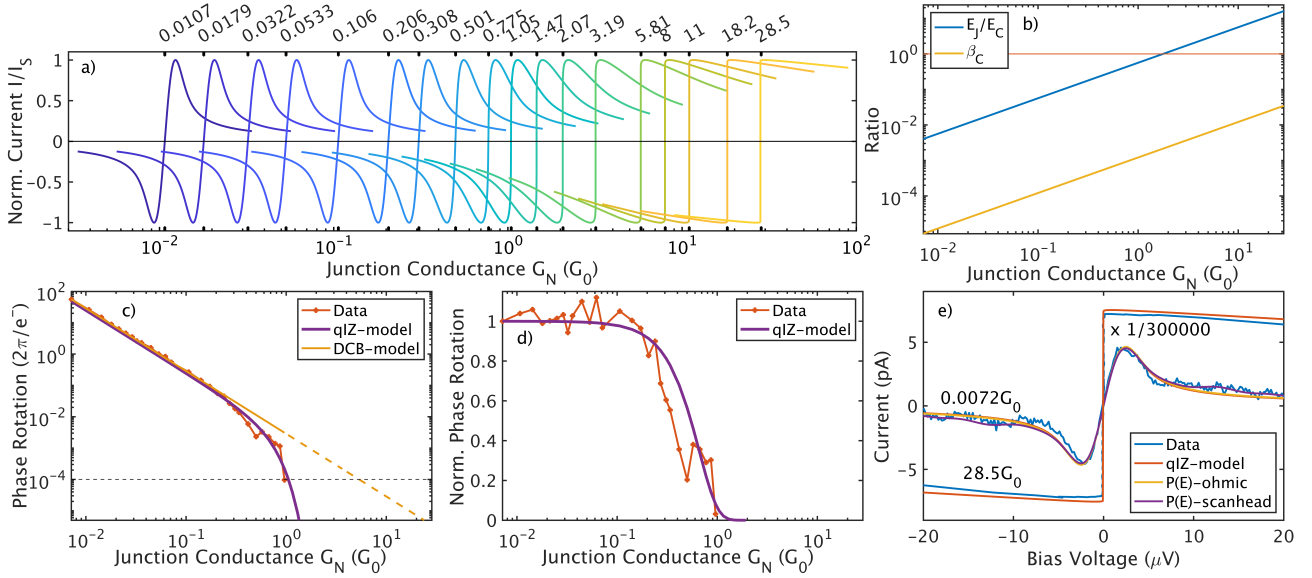}
\caption{\label{Fig4}\panelcaption{a} Evolution of the Josephson current of the qIZ model within a voltage window of $\pm 20\,\upmu$V as function of $G_\text{N}$ for the same setpoints as in \figref[c]{Fig1}. The change in the slope at zero voltage and the reduced interaction with the environment follow the same evolution as in the experimental data. \panelcaption{b} Ratio $E_\mathrm{J}/E_\mathrm{C}$ and Stewart-McCumber parameter $\beta_\text{C}$ as function of $G_\text{N}$. The transition region where $E_\mathrm{J}/E_\mathrm{C}\approx 1$ lies around $G_0$, in agreement with the experimentally observed crossover. The Stewart-McCumber parameter remains below one over the full range, $\beta_\text{C}<1$, so that the junction is always in the overdamped regime. \panelcaption{c} Phase rotation per tunneling Cooper pair $\frac{\Delta\phi}{\Delta n}$ as function of $G_\text{N}$ for the experimental data (red line with markers), the qIZ model (purple line), and the DCB model (yellow line). The phase rotation slows down in the transition regime, deviating from the DCB prediction. The horizontal dashed line marks the sensitivity limit of the mK-STM. \panelcaption{d} Phase rotation per tunneling Cooper pair normalized to the value, which is expected from the DCB. In the DCB, the phase rotation depends quadratically on the junction conductance $\frac{\Delta\phi}{\Delta n}\big|_\text{DCB}\propto G_\text{N}^2$. The first indications of a slowing phase can be seen around $0.1\,G_0$ already. \panelcaption{e} Comparison of the experimental $I(V)$-curves with the qIZ model fits for the two extreme values of $G_\text{N}$, together with two other models ($P(E)$-theory using an ohmic impedance, and using the impedance model of the scan head~\cite{zeng_enhancing_2026}). The agreement is excellent, particularly in the high $G_\text{N}$ regime, given the simplifying assumptions of the qIZ model.}
\end{figure*}

The model also provides the Stewart-McCumber parameter $\beta_\text{C}$ as a function of $G_\text{N}$ (\figref[b]{Fig4}), which in a conventional junction distinguishes a hysteretic, underdamped~\cite{goffman_supercurrent_2000} ($\beta_\text{C}>1$) from a non-hysteretic, overdamped ($\beta_\text{C}<1$) junction. It predicts that $E_\mathrm{J}/E_\mathrm{C}\approx 1$ around $G_0$, precisely in the transition regime where we observe the quadratic-to-linear crossover, while $\beta_\text{C}$ remains well below one over the entire range. The junction is thus always in the overdamped regime, a crucial assumption of the qIZ model~\cite{ankerhold_overdamped_2004}.

The accuracy of the data also allows us to follow how the charge-phase complementarity evolves as we sweep through the crossover region. The uncertainty product of charge and phase is not directly accessible in the present experiment. What we can access is the phase rotation per tunneling Cooper pair $\Delta\phi/\Delta n$, which serves as a transport indicator of where the junction sits between the charge- and phase-dominated regimes: the tunneling current and the voltage measure the change in particle number and in phase per unit time, respectively, from which we derive $\Delta\phi/\Delta n = 4\pi\, G_0/G_\text{ZB}$ (derivation in the SM). \Figref[c]{Fig4} shows this quantity from the experimental data (red line with markers) together with the qIZ (purple line) and DCB (yellow line) predictions; the horizontal dashed line marks the sensitivity limit of the mK-STM. Around $G_\text{N}\approx G_0$, $\Delta\phi/\Delta n$ decreases faster than the DCB prediction, so the phase changes much less with increasing current and eventually becomes constant, i.e.\ independent of the current, as expected in the Josephson (DCT) regime. \Figref[d]{Fig4} shows the same data on a linear scale, normalized to the value expected from the DCB model, making explicit how the phase slows down and decouples, which also means that $G_\text{ZB}$ becomes independent of $G_\text{N}$.

While phase fluctuations are strongly suppressed at high conductance ($G_\text{N}\gg G_0$, Josephson regime), the opposite holds at low conductance ($G_\text{N}\ll G_0$), where they far exceed $2\pi$ and cover many periods of the classical Josephson potential $V_\text{J}(\phi)=-E_\text{J} \cos(\phi)$ and its generalization to few transmission channels. The semiclassical picture of a phase trajectory surrounded by weak quantum fluctuations then becomes meaningless, and the charge at the junction turns into the appropriate and well-defined degree of freedom. This is the domain of the DCB, where the exchange of energy quanta with the electromagnetic environment governs the Cooper pair current via the $P(E)$-function, whose essential ingredient is the phase-phase correlator.

Finally, \figref[e]{Fig4} compares the experimental $I(V)$-curves with the qIZ model at the two extreme setpoints, using identical parameters with only the Josephson energy $E_\text{J}$ adjusted to match $G_\text{N}$. The agreement is excellent in both cases, even though the tunneling currents differ by more than five orders of magnitude. For the low-conductance spectrum, we also show the Josephson current from $P(E)$-theory with an ohmic impedance (yellow line), which matches the qIZ model as expected, and from the more realistic scan-head impedance~\cite{zeng_enhancing_2026}, which reproduces the environmental resonances seen in the data. The high-conductance curve, whose parameters were fixed at the low-conductance spectrum, agrees very well in its overall shape. Despite its simplifying assumptions, the qIZ model captures the evolution across the quadratic-to-linear transition and provides a consistent picture across all observables.

In conclusion, we have observed the transition between two seemingly disconnected regimes of Josephson tunneling -- the dynamical Coulomb blockade and dissipationless Cooper pair tunneling regimes -- in a single tunnel junction. The transition is smooth and continuous with an increasing number of transport channels and is reasonably reproduced by the qIZ model within its simplifying assumptions. These results also confirm that the dependence of the Cooper pair current on the electromagnetic environment fades away towards the phase-dominated regime. Several aspects warrant further theoretical attention, in particular the treatment of an arbitrary rather than purely ohmic electromagnetic environment, and the inclusion of higher harmonics in the current-phase relation when high-transmission channels are present. Linking dissipative and dissipationless Cooper pair tunneling in this way also bridges the few-channel STM tunnel junction, with its sequential charge tunneling, and the many-channel planar tunnel junction, with its coherent phase tunneling. Continuously tuning the complementarity between two conjugate quantum-mechanical variables in a single device opens new avenues for systematic studies of the quantum-to-classical transition in superconducting circuits.

\section*{Acknowledgments}
We gratefully acknowledge stimulating discussions with Björn Kubala, Ciprian Padurariu. This work was funded in part by the Center for Integrated Quantum Science and Technology (IQ$^\textrm{\small ST}$). J.C.C.\ thanks the Spanish Ministry of Science and Innovation (contract no.\ PID2024-157536NB-C22) and the ``Maria de Maeztu'' Programme for Units of Excellence in R\&D (CEX2023-001316-M).

\section*{Data Availability}
All data needed to evaluate the conclusions are present in this paper and/or the Supplementary Material.

\section*{Code Availability}
The code used for the calculations presented here is available from the authors upon reasonable request.

\clearpage
\newpage

\onecolumngrid
\begin{center}
\textbf{\large Supplementary Material}
\end{center}
\vspace{1cm}
\twocolumngrid

\setcounter{secnumdepth}{1}

\makeatletter\global\let\@FMN@list\@empty\makeatother

\setcounter{figure}{0}
\setcounter{table}{0}
\setcounter{equation}{0}
\renewcommand{\thefigure}{S\arabic{figure}}
\renewcommand{\thetable}{S\Roman{table}}
\renewcommand{\theequation}{S\arabic{equation}}

\section{Methods}

The measurements were performed using a home-built scanning tunneling microscope (STM) attached to a dilution refrigerator that reaches a base temperature of 10\,mK in the mixing chamber~\cite{assig_10_2013_si}. The STM is operated in ultra-high vacuum and offers \textit{in situ} sample preparation. To improve the energy resolution, the STM is equipped with a local electromagnetic shielding combined with low-pass filtering at the cryogenic scan head, which suppresses the high-frequency radiation reaching the junction and capacitively shunts the tunnel junction~\cite{zeng_enhancing_2026_si}. 

The tip is a polycrystalline vanadium wire that is mechanically cut and conditioned \textit{in situ} by controlled indentation into the sample. The sample is a V(100) single crystal cleaned by repeated sputtering and annealing cycles. Both tip and sample become superconducting below $T_\text{c}\approx 5.4\,$K with a gap parameter in the sample of $\Delta\approx 750\,\upmu$eV. Due to imperfections in the tip, the gap parameter in the tip is $\Delta\approx 700\,\upmu$eV.

The tunneling current $I(V)$ was acquired as function of applied voltage (voltage bias) in constant-height mode for a fixed tip-sample distance. The setpoint current $I_\text{set}$ was changed by varying the tip-sample distance at a fixed reference voltage $V_\text{set}=4\,$mV. To capture the multiple Andreev reflection (MAR) features, $I(V)$ curves were recorded over the full range $|V|<4\,$mV at each setpoint. To capture the Josephson current, the point density was increased in a small voltage window near zero with sub-microvolt resolution. 

\section{Data Analysis}

The junction voltage $V_\text{J}$ was calculated from the applied bias voltage $V_\text{B}$ by correcting for the voltage drop across the line resistance $R_\mathrm{L}=484.1\,\upOmega$ using
\begin{equation}
    V_\text{J} = V_\text{B} - R_\mathrm{L}\cdot I(V_\text{B}).
\end{equation}

The switching current $I_\text{S}$ was extracted as the maximum of the Josephson current within the narrow-bias window, taken from the average of the absolute values at positive and negative bias to suppress any residual offset. The zero-bias conductance $G_\text{ZB}$ was extracted from a linear fit to the current-voltage characteristic over a $\pm 1.3\,\upmu$V window centered at zero. The junction conductance values reported in the main text and figures are the normal state conductance $G_\text{N}$ near the setpoint voltage outside of the superconducting gap.

\section{Additional Datasets}

\begin{figure*}
\centering\includegraphics[width=1\textwidth]{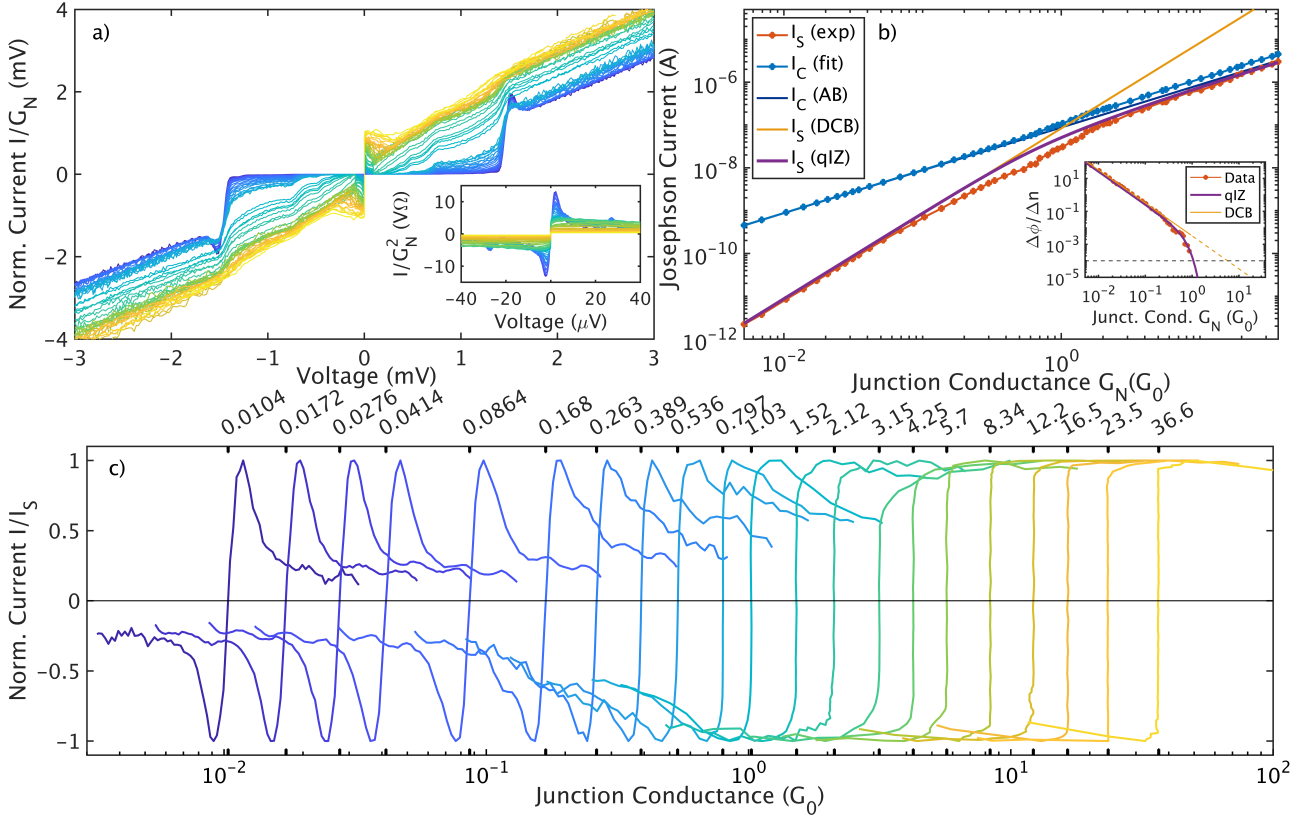}
\caption{\label{FigS1} \panelcaption{a} Tunneling current vs.\ voltage for different values of the junction conductance $0.007\,G_0\leq G_\text{N} \leq 36.6\,G_0$, normalized to $G_\text{N}$ for another dataset. The coherence peaks are visible at $\pm1.46\,$mV. The offset in the normal-state regime is due to the excess current; the steps inside the gap are from multiple Andreev reflections. The Josephson effect at high conductance is visible near zero voltage. The lower inset shows the tunneling current normalized to $G_\text{N}^2$ to highlight the low-conductance Josephson effect. \panelcaption{b} Josephson current as function of junction conductance. The critical current $I_\text{C}$ evolves linearly and is calculated both from the Ambegaokar-Baratoff (AB) formula (blue line) and from the channel-resolved fit to the experimental data (blue line with markers). The switching current $I_\text{S}$ initially evolves quadratically and then linearly, both for the experimental data (red line with markers) and for the qIZ model (purple line). $I_\text{S}$ has been extracted from the $I(V)$-curves as illustrated in the upper inset. The yellow line shows the quadratic evolution expected from the DCB model. The experimental $I_\text{S}$ is in good agreement with the qIZ model in the extreme limits; in the transition regime, deviations are visible that may be traced to the simplifying assumptions of the model. The lower inset shows the ratio $\Delta\phi/\Delta n$, which behaves very similarly to the dataset in the main text (cf.\ Fig.\ 4(c)). \panelcaption{c} Evolution of the Josephson current within a voltage window of $\pm 20\,\upmu$V as function of $G_\text{N}$ for selected $I(V)$-curves from \panelsubcaption{a}. The change in the slope at zero voltage and the reduced interaction with the environment are clearly visible.}
\end{figure*}

In \figref{FigS1}, another dataset is shown for a different tunnel junction using the same materials as in the main text (V tip, V(100) sample). The set of $I(V)$-curves is shown in \figref[a]{FigS1} with the inset zooming in to the Josephson current. In \figref[b]{FigS1}, the evolution of the Josephson current maximum is shown in comparison to the evolution of the different models discussed in the main text. We see an overall agreement with the models in their respective range of validity similar to the dataset in the main text. The qIZ-model calculation is the same as in the main text. In \figref[c]{FigS1}, the evolution of the Josephson current is shown again displaying the same behavior as the dataset in the main text. This demonstrates that the general behavior is not specific to a particular tunnel junction, but exhibits a more general behavior. This also includes the deviation of the experimental data from the qIZ-model in the vicinity of $G_0$. 

\section{Fitting the Transport Channels}
\phantomsection\label{sec:channel_fits}

\subsection{Theoretical framework}

The number of transport channels and their individual transmissions are extracted by fitting the quasiparticle $I(V)$ curves outside the Josephson region to the theory of a voltage-biased superconducting point contact~\cite{cuevas_hamiltonian_1996_si}. The formalism provides a microscopic, nonperturbative description of the charge transport between two superconducting electrodes connected by a single quantum channel of normal transmission~$\tau$. The key feature of this approach is that the coupling between the electrodes is treated to all orders, thereby capturing the full subgap current together with the gap-edge and excess current features outside of the gap in a unified framework, in agreement with the corresponding scattering-theory results~\cite{cuevas_hamiltonian_1996_si}.

\paragraph{Hamiltonian model and Current formula.} The starting point is a tight-binding model in which each superconducting electrode is described by a BCS Hamiltonian and the two electrodes are coupled by a single hopping element across the contact. The model Hamiltonian reads
\begin{equation}
\hat{H}=\hat{H}_\text{L}+\hat{H}_\text{R}+\sum_\sigma\!\left(t\,c^\dagger_{\text{L}\sigma}c_{\text{R}\sigma}+t^*\,c^\dagger_{\text{R}\sigma}c_{\text{L}\sigma}\right)-\mu_\text{L}\hat{N}_\text{L}-\mu_\text{R}\hat{N}_\text{R},
\label{eq:Hcontact}
\end{equation}
where $\hat{H}_\text{L,R}$ are the BCS Hamiltonians of the uncoupled electrodes, $c^{(\dagger)}_{\text{L/R}\sigma}$ annihilates (creates) an electron of spin $\sigma$ on the outermost site of the left or right electrode, $\hat{N}_\text{L,R}$ are the corresponding electron number operators, $t$ is the hopping amplitude, and the chemical potentials are fixed by the applied bias through $eV=\mu_\text{L}-\mu_\text{R}$. Despite its simplicity, the model captures the relevant physics of a single-channel superconducting point contact, with the normal transmission~$\tau$ tunable continuously from the tunnel limit ($\tau\ll 1$) to the ballistic limit ($\tau=1$) by varying the hopping parameter $t$~\cite{cuevas_hamiltonian_1996_si}. The normal transmission~$\tau$ of one channel is related to the microscopic hopping~$t$ in the small-bias voltage limit by~\cite{cuevas_hamiltonian_1996_si}
\begin{equation}
\tau=\frac{4\tilde{t}^{\,2}}{(1+\tilde{t}^{\,2})^{2}},
\label{eq:tau_t_rho}
\end{equation}
where $\tilde{t}^2\equiv \pi^2 t^2\rho_{\text{L}}\,\rho_{\text{R}}$ and $\rho_{\text{L}}$ and $\rho_{\text{R}}$ are the normal-state densities of states at the Fermi level in the left and right electrodes, respectively, taken to be constant over the relevant bias voltage range. Equation~\eqref{eq:tau_t_rho} is a nonperturbative relation valid for arbitrary $t$. In the normal state the model reproduces the Landauer formula $G=G_0\tau$, with $G_0=2e^2/h$.

\paragraph{Harmonic index and matrix structure.} Under an applied bias voltage, the time-dependent superconducting phase difference $\phi(t)=\phi_0+2eVt/\hbar$, with $\phi_0$ the static phase offset, is absorbed into a phase factor on the hopping element, $t\to t\,\mathrm{e}^{i\phi(t)/2}$, so that the problem becomes explicitly time-periodic with fundamental frequency $\omega_0=2eV/\hbar$. The current
\begin{equation}
I(t)=\frac{ie}{\hbar}\sum_\sigma\!\left[t\,\langle c^\dagger_{\text{L}\sigma}(t)\,c_{\text{R}\sigma}(t)\rangle - t^*\,\langle c^\dagger_{\text{R}\sigma}(t)\,c_{\text{L}\sigma}(t)\rangle\right]
\label{eq:Ifrom_corr}
\end{equation}
can therefore be expanded as a Fourier series in harmonics of $\omega_0$,
\begin{equation}
I(t)=\sum_{m=-\infty}^{\infty} I_m\,\mathrm{e}^{im\omega_0 t}.
\label{eq:Ifourier}
\end{equation}
The relevant correlation functions are obtained from the nonequilibrium Green's functions of the coupled system, which are computed in 2$\times$2 Nambu space by solving the Dyson equation for the dressed propagators with the hopping treated as a selfenergy. The result is valid for arbitrary $\tau$ across the full bias voltage range~\cite{cuevas_hamiltonian_1996_si}.

The quantity entering the fits is the dc component~$I_0(V)\equiv I_\tau(V)$ of Eq.~\eqref{eq:Ifourier}. It is a single, smooth function of bias and transmission, computed numerically by solving the recursion relations for the renormalized hopping coefficients of reference~\cite{cuevas_hamiltonian_1996_si}. The full nonlinear $\tau$-dependence of $I_\tau(V)$, including the subharmonic gap structure (SGS) at $eV\sim 2\Delta/n$ and the smooth crossover to the excess current behavior outside the gap, is what makes the spectrum a sensitive probe of the number of transport channels and their transmission.

We label the Fourier components by an integer index $n$, organize them along the diagonal of an infinite block matrix, and treat the resulting object as the fundamental quantity of the formalism. With this convention the unperturbed (retarded/advanced) Green's function of electrode $\alpha=\text{L,R}$ is block-diagonal,
\begin{align}
    \hat{g}^{r,a}_{\alpha}(\omega)&=-\frac{1}{\sqrt{\Delta_\alpha^2-(\omega\mp i\eta)^2}}\begin{pmatrix}\omega\mp i\eta & \Delta_\alpha\\ \Delta_\alpha & \omega\mp i\eta\end{pmatrix},\label{eq:g_block}\\
    \hat{g}^{r,a}_{\alpha;nm}(\omega)&=\delta_{nm}\,\hat{g}^{r,a}_{\alpha}(\omega+neV),
\end{align}
each $2\times 2$ block being the standard Nambu BCS Green's function shifted by $n$ harmonics, with $\Delta_\alpha$ being the gap of electrode $\alpha$ and $\eta>0$ a small (Dynes-type) inelastic broadening. The lesser component carrying the Fermi distribution follows from the fluctuation-dissipation relation, 
\begin{equation}
    \hat{g}^{+-}_{\alpha;nm}=\delta_{nm}\,[\hat{g}^{a}_{\alpha}-\hat{g}^{r}_{\alpha}](\omega+neV)\,n_\text{F}(\omega+neV),
\end{equation}
where $n_\text{F}(\omega)$ is the Fermi function at temperature~$T$.

The hopping is block-tridiagonal. In the rotating gauge an electron-like hop carries one phase factor $\mathrm{e}^{+i\phi/2}$, which shifts the harmonic index by $+1$, and a hole-like hop carries $\mathrm{e}^{-i\phi/2}$, shifting by $-1$. Explicitly,
\begin{equation}
\hat{t}_{LR;n,n+1}=\begin{pmatrix}t & 0\\ 0 & 0\end{pmatrix},\qquad
\hat{t}_{LR;n,n-1}=\begin{pmatrix}0 & 0\\ 0 & -t\end{pmatrix},\qquad
\hat{t}_{RL}=\hat{t}_{LR}^{\dagger},
\label{eq:t_block}
\end{equation}
with all other blocks vanishing. Equations~\eqref{eq:g_block} and~\eqref{eq:t_block} encode the entire harmonic index structure of the bias voltage driven problem. The gap structure of both electrodes lives in $\hat g$, while the bias-induced harmonic shifts and the Nambu (electron/hole) structure live in $\hat t$.

\paragraph{Dressed hopping and all-orders resummation.} Treating $\hat t$ as a self-energy in the Dyson equation for the coupled propagators leads to a closed integral equation for the renormalized hopping $\hat T^{r,a}_{LR}$, which accounts for all repeated hops across the junction~\cite{cuevas_hamiltonian_1996_si}. In the block notation introduced above this equation takes the compact form
\begin{equation}
\hat{T}^{r,a}_{LR}=\hat{t}_{LR}+\hat{Q}^{r,a}_{LL}\,\hat{T}^{r,a}_{LR},
\qquad
\hat{Q}^{r,a}_{LL}\equiv\hat{t}_{LR}\,\hat{g}^{r,a}_{RR}\,\hat{t}_{RL}\,\hat{g}^{r,a}_{LL},
\label{eq:Tsolve}
\end{equation}
with formal solution
\begin{equation}
\hat{T}^{r,a}_{LR}=\frac{1}{1-\hat{Q}^{r,a}_{LL}}\,\hat{t}_{LR}=\sum_{k=0}^{\infty}\!\left(\hat{Q}^{r,a}_{LL}\right)^{k}\!\hat{t}_{LR}.
\label{eq:Tseries}
\end{equation}
The geometric series in Eq.~\eqref{eq:Tseries} is the all-orders resummation of the hopping processes, and the resulting $\hat T$ is the only object in the formalism that depends on the contact transmission once the BCS Green's functions are fixed by $\Delta_\text{L}$ and $\Delta_\text{R}$.

\paragraph{Current.} The dc current is a bilinear functional of $\hat T$ in the harmonic-index space~\cite{cuevas_hamiltonian_1996_si},
\begin{equation}
\begin{aligned}
I_0(V&) = I_\tau(V) =\frac{e}{h}\!\int_{-\infty}^{\infty}\!\!d\omega\sum_{n\,\text{odd}}\mathrm{Tr}\Big\{\hat{\sigma}_z
\Big[\\
&\hat{T}^{r}_{LR,0n}\,\hat{g}^{+-}_{R,nn}\,\hat{T}^{a}_{RL,n0}\,\hat{g}^{a}_{L,00}
+\hat{g}^{r}_{R,00}\,\hat{T}^{r}_{RL,0n}\,\hat{g}^{+-}_{L,nn}\,\hat{T}^{a}_{LR,n0}\\
&\!\!\!\!\!\!-\hat{g}^{r}_{L,00}\,\hat{T}^{r}_{LR,0n}\,\hat{g}^{+-}_{R,nn}\,\hat{T}^{a}_{RL,n0}
-\hat{T}^{r}_{RL,0n}\,\hat{g}^{+-}_{L,nn}\,\hat{T}^{a}_{LR,n0}\,\hat{g}^{a}_{R,00}\Big]\Big\},
\end{aligned}
\label{eq:I0_dc}
\end{equation}
where $\hat\sigma_z=\mathrm{diag}(1,-1)$ is the third Pauli matrix in Nambu space, ensuring the correct relative sign between the electron and hole contributions to the current. The trace runs over the $2\times 2$ Nambu indices, and the sum runs over odd harmonics.

\paragraph{Multichannel decomposition.} In a junction having several transport channels with transmission $\{\tau_i\}$, the channels are independent and non-interacting~\cite{cuevas_hamiltonian_1996_si,scheer_conduction_1997_si}, so that the total quasiparticle current is the sum of the individual channel contributions
\begin{equation}
I(V)=\sum_i I_{\tau_i}(V),
\label{eq:Imulti}
\end{equation}
The total normal state conductance is calculated as
\begin{equation}
    G_\text{N}=G_0\sum_i\tau_i,
\end{equation}
where $G_0=2e^2/h$ is the quantum of conductance. A given $G_\text{N}$ can be realized by very different channel sets $\{\tau_i\}$, and these alternatives produce markedly different SGS amplitudes because $I_{\tau_i}(V)$ depends nonlinearly on $\tau_i$ through the resummation in Eq.~\eqref{eq:Tseries}. A fit to the full subgap IV therefore constrains the set $\{\tau_i\}$ uniquely, well beyond what the total conductance alone provides~\cite{scheer_conduction_1997_si}.

\subsection{Asymmetric STM junction}

In our experiment both electrodes are vanadium, with a measured gap parameter $\Delta\approx 750\,\upmu$eV in the ideal case. In practice, however, the tip apex is shaped by controlled indentation into the sample, which can leave the tip and sample with slightly different effective gaps, $\Delta_\text{tip}$ and $\Delta_\text{sample}$. In the Green's-function derivation above the two electrodes are labelled L and R; for the STM junction we identify the left electrode with the tip and the right with the sample, $\mathrm{L}\equiv\mathrm{tip}$ and $\mathrm{R}\equiv\mathrm{sample}$, so that $\Delta_\mathrm{L}=\Delta_\mathrm{tip}$ and $\Delta_\mathrm{R}=\Delta_\mathrm{sample}$. The single-channel current $I_\tau(V)$ used in the fits is computed for the corresponding asymmetric S-S$'$ junction, which produces additional features in the SGS at bias voltages of the form $eV=(\Delta_\text{tip}+\Delta_\text{sample})/n$ and $eV=|\Delta_\text{tip}\pm\Delta_\text{sample}|/n$~\cite{scheer_conduction_1997_si,ternes_subgap_2006_si}. The largest feature is the quasiparticle onset at $eV=\Delta_\text{tip}+\Delta_\text{sample}$, whose position is read off directly from the coherence peaks. The values of $\Delta_\text{tip}$ and $\Delta_\text{sample}$ are extracted independently from the low-bias coherence-peak structure of each spectrum and held fixed during the channel fit, so that the only free parameters are the transmissions $\{\tau_i\}$.

\subsection{Fitting procedure}

Following the procedure established by Scheer~\textit{et al.}~\cite{scheer_conduction_1997_si}, Eq.~\eqref{eq:Imulti} is fitted to the experimental quasiparticle spectrum $I(V)$ for each setpoint, using the channel transmissions $\{\tau_i\}$ as free parameters and $\Delta_\text{tip}$, $\Delta_\text{sample}$ fixed at the values determined from the coherence peaks. The numerical single-channel current $I_\tau(V)$ is precomputed on a dense grid in $\tau$ and $V$ from the Hamiltonian formalism of reference~\cite{cuevas_hamiltonian_1996_si}, and the fit minimizes the squared residuals.

To avoid overfitting, channels are added one at a time: starting from a single channel, an additional channel is included only when its inclusion produces a statistically significant reduction in the residuals~\cite{scheer_conduction_1997_si}. The final number of channels obtained in this way ranges from a single channel of low transmission at the lowest conductances ($G_\text{N}\lesssim 0.1\,G_0$) to up to 42 channels at the highest conductance ($G_\text{N}\approx 28.5\,G_0$).

The accuracy with which individual transmissions can be extracted depends on the number of contributing channels and on their absolute values, and follows the analysis of reference~\cite{scheer_conduction_1997_si}. For $N\leq 3$ contributing channels, each $\tau_i$ is determined to within $\sim 1\%$ of the total conductance $G_\text{N}$. For larger $N$, only the few dominant channels are determined to that accuracy, while the smaller transmissions become less well constrained. The reason is that the information that disentangles additional or low-transmission channels is contained in the deep-subgap part of the spectrum, where the current amplitude is smallest. Crucially, the total number of channels with $\tau_i$ above the $\sim 1\%$ threshold and the total conductance $G_\text{N}$ is robustly determined by the fit, even when individual small transmissions are uncertain~\cite{scheer_conduction_1997_si}.

To verify whether a single-channel description suffices, we have performed the channel fits both with a single channel and with the multichannel ansatz of Eq.~\eqref{eq:Imulti}. For low conductances, $G_\text{N}\lesssim 0.1\,G_0$, a single channel of low transmission reproduces the spectrum to within the noise, consistent with the expectation that the conductance is dominated by tunneling through a single dominant channel. For $G_\text{N}\gtrsim G_0$ the single-channel fit fails: the SGS amplitudes cannot be reproduced by any single transmission, and several channels with different transmissions are required. As $G_\text{N}$ is increased further, additional channels open and channels of high transparency, $\tau\approx 0.96\pm 0.03$, accumulate. At the highest conductance the junction has up to 42 channels, several of them in the high-transparency regime. The resulting channel statistics is consistent with that of mesoscopic atomic-size contacts of $s$-$d$ metals~\cite{scheer_conduction_1997_si,scheer_signature_1998_si}.

\section{Multichannel Josephson Junctions} \label{sec:multichannel_josephson}

The Josephson coupling of the multichannel STM junction is computed within the same nonequilibrium Green's-function framework as the quasiparticle current of Section~\ref{sec:channel_fits}. We follow the microscopic treatment in Ref.~\cite{martin-rodero_microscopic_1994_si}, which yields a closed expression for the current-phase relation of a ultrasmall single-channel $SIS'$ contact directly in terms of the uncoupled-electrode Green's functions $\hat g^r_\text{L}(\omega)$ and $\hat g^r_\text{R}(\omega)$ introduced in Eq.~\eqref{eq:g_block}. Because both electrodes of our STM junction are vanadium but the tip and the sample are reshaped by repeated indentation, the effective gaps $\Delta_\text{tip}$ and $\Delta_\text{sample}$ generally differ, so an asymmetric formulation is required.

\paragraph{Single-channel current-phase relation.} In the short-junction limit ($L_\mathrm{C}\ll\xi_0$), where $L_\mathrm{C}$ is the length of the contact region and $\xi_0$ the superconducting coherence length. The order-parameter profile may be approximated by a step function across the contact, $\phi=\phi_\text{L}-\phi_\text{R}$, and the supercurrent through a single channel of normal transmission $\tau$ can be written as~\cite{martin-rodero_microscopic_1994_si}
\begin{equation}
I_{\text{S},\tau}(\phi)=\frac{8e}{\hbar}\,t^2\,\sin\phi\int_{-\infty}^{\infty}\!\!d\omega\,\mathrm{Im}\!\left[\frac{\tilde g^{\,r}_{\text{L},21}(\omega)\,\tilde g^{\,r}_{\text{R},12}(\omega)}{D^r(\omega)}\right]\!n_\text{F}(\omega),
\label{eq:Iphi_MR}
\end{equation}
with
\begin{equation}
D^r(\omega)=\det\!\left[\hat 1-t^2\,\hat\sigma_z\,\hat g^{\,r}_\text{L}(\omega)\,\hat\sigma_z\,\hat g^{\,r}_\text{R}(\omega)\right].
\label{eq:Dr_MR}
\end{equation}
Here $\hat g^{\,r}_{\alpha}(\omega)$ are the retarded Nambu Green's functions of the uncoupled electrodes given in Eq.~\eqref{eq:g_block}, the tilde indicates that the static phases have been gauged out ($ g^{\,r}_{\text{L},21}=\mathrm{e}^{i\phi_\text{L}}\tilde g^{\,r}_{\text{L},21}$, $ g^{\,r}_{\text{R},12}=\mathrm{e}^{-i\phi_\text{R}}\tilde g^{\,r}_{\text{R},12}$), $n_\text{F}(\omega)$ is the Fermi function at temperature~$T$, and $t$ is the hopping parameter across the contact, related to the normal-state transmission $\tau$ as shown in Eq.~\eqref{eq:tau_t_rho}. The denominator $D^r(\omega)$ encodes the resummation of all higher order processes.

\paragraph{Multichannel sum.} For a junction with several independent transport channels of transmissions $\{\tau_i\}$, the total supercurrent is the sum of the single-channel contributions
\begin{equation}
I_\text{S}(\phi)=\sum_i I_{\text{S},\tau_i}(\phi),
\label{eq:Iphi_total}
\end{equation}
where each $I_{\text{S},\tau_i}$ is computed from Eq.~\eqref{eq:Iphi_MR} using the same $\Delta_\text{tip}$ and $\Delta_\text{sample}$ extracted from the coherence peaks of the quasiparticle spectrum (see Section~\ref{sec:channel_fits}). The transmissions $\{\tau_i\}$ are those obtained from the MAR channel fit. The corresponding hopping values $t$ can be calculated from the inverse of Eq.~\eqref{eq:tau_t_rho}.
Equation~\eqref{eq:tau_t_rho} can be inverted to express the dimensionless coupling in terms of the transmission
\begin{equation}
t^{\,2}=\frac{1}{\pi^2\rho_{\text{L}}\,\rho_{\text{R}}}\frac{1-\sqrt{1-\tau}}{1+\sqrt{1-\tau}}.
\label{eq:t_of_tau}
\end{equation}
The critical current $I_\text{C}$ is then the maximum of $I_\text{S}(\phi)$ over $\phi\in[0,2\pi]$.

In the tunneling limit $\tau_i\ll 1$, Eq.~\eqref{eq:Iphi_MR} reduces to a sinusoidal current-phase relation $I_{\text{S},i}\propto\tau_i\sin\phi$, and the sum gives the Ambegaokar-Baratoff result generalized to unequal gaps~\cite{ambegaokar_tunneling_1963_si,golubov_current_2004_si,barone_physics_1982_si},
\begin{equation}
I_\text{C}^\text{AB}=\frac{G_\text{N}}{e}\,\frac{2\Delta_\text{tip}\Delta_\text{sample}}{\Delta_\text{tip}+\Delta_\text{sample}}\,K\!\left(\frac{|\Delta_\text{tip}-\Delta_\text{sample}|}{\Delta_\text{tip}+\Delta_\text{sample}}\right),
\label{eq:IcAB_asym}
\end{equation}
with $K$ being the complete elliptic integral of the first kind, which reduces to $I_\text{C}^\text{AB}=\pi\Delta\,G_\text{N}/(2e)$ when $\Delta_\text{tip}=\Delta_\text{sample}=\Delta$. As channels with high transmission are added, the current-phase relation deviates from a sinusoid and $I_\text{C}$ exceeds the Ambegaokar-Baratoff value.

The current-phase relation $I_\text{S}(\phi)$ shown in Fig.~3(c) of the main text is computed for each setpoint by inserting the fitted channel transmissions $\{\tau_i\}$ into Eqs.~\eqref{eq:Iphi_MR} and~\eqref{eq:Iphi_total} at $T=10\,$mK, using the experimentally determined gaps $\Delta_\text{L}$ and $\Delta_\text{R}$. In the tunneling limit, the current-phase relation is sinusoidal, $I_\text{S}\propto\sin\phi$. With increasing transmission, the maximum shifts toward $\phi>\pi/2$ and the relation acquires higher harmonics. For a fully transparent channel ($\tau=1$), the zero-temperature current-phase relation reduces to $I_\text{S}\propto\sin(\phi/2)$ for $\phi\in[0,2\pi)$, with a sawtooth-like jump at $\phi=2\pi$. None of our setpoints reaches the fully transparent limit, so the deviation from a sinusoid in our experiment, although noticeable, remains smooth and continuous, as visible in Fig.~3(c) of the main text.

\section{Theoretical Model at Arbitrary Coupling}
\label{sec:qIZ}

The theoretical model used in the main text to describe the Cooper pair current across the full crossover from the dynamical Coulomb blockade (DCB) regime to the dissipationless Cooper pair tunneling (DCT) regime is taken from Grabert, Ingold, and Paul~\cite{grabert_phase_1998_si}, which we refer to in the main text as the quantum Ivanchenko-Zil'berman (qIZ) model~\cite{ivanchenko_josephson_1969_si}. We summarize here the derivation and the assumptions on which it relies, as both are relevant for the comparison with the experiment in Figs.~2 and~4 of the main text.

\subsection{Hamiltonian and assumptions}

We consider a Josephson junction with capacitance $C_\mathrm{J}$ and Josephson energy $E_\mathrm{J} = (\hbar/2e)\,I_\mathrm{C}$, voltage-biased through an external resistance~$R$ that represents its electromagnetic environment. The total Hamiltonian can be written as the sum of a junction part and an environmental part,
\begin{equation}
H = H_\mathrm{junction} + H_\mathrm{env}.
\label{eq:Htotal}
\end{equation}
The junction Hamiltonian contains the Cooper-pair charging energy and the Josephson coupling~\cite{josephson_possible_1962_si,tinkham_introduction_1996_si},
\begin{equation}
H_\mathrm{junction} = E_\mathrm{C}\,\hat{n}^{2} - E_\mathrm{J}\cos\hat{\varphi},
\label{eq:Hjunction}
\end{equation}
where the Cooper-pair charging energy is $E_\mathrm{C}=2e^{2}/C_\mathrm{J}$, the Cooper-pair number operator $\hat{n}=\hat{Q}/2e$ counts the charges transferred across the junction in units of $2e$, and $\hat{\varphi}$ is the superconducting phase difference. The two operators are conjugate,
\begin{equation}
[\hat{\varphi},\hat{n}] = i,
\label{eq:commutator}
\end{equation}
which is the form used throughout the main text. Equation~\eqref{eq:Hjunction} makes the competition between the two relevant energy scales explicit: when $E_\mathrm{J}\gg E_\mathrm{C}$, the cosine potential dominates and the phase $\hat{\varphi}$ is well defined; when $E_\mathrm{J}\ll E_\mathrm{C}$, the kinetic term dominates and the number $\hat{n}$ is well defined.

The environmental part couples the junction phase to a bath of $LC$-oscillators that represents the external resistor~\cite{caldeira_quantum_1983_si,ingold_charge_1992_si,vool_introduction_2017_si},
\begin{equation}
H_\mathrm{env} = \sum_{n=1}^{\infty}\!\left[\frac{\hat{q}_n^{2}}{2C_n}
+ \!\left(\frac{\hbar}{2e}\right)^{\!2}\!\frac{1}{2L_n}(\hat{\varphi}_\mathrm{R}-\hat{\varphi}_n)^{2}\right]\!,
\label{eq:Henv}
\end{equation}
where $\hat{\varphi}_\mathrm{R}=(2e/\hbar)Vt-\hat{\varphi}$ is the phase associated with the voltage drop across the resistor, $\hat{q}_n$ and $\hat{\varphi}_n$ are the charge and phase coordinates of the $n$-th bath mode, and the parameters $C_n$, $L_n$ are chosen such that the linear response of the bath reproduces the impedance of the external $RC$ circuit~\cite{caldeira_quantum_1983_si,ingold_charge_1992_si,grabert_phase_1998_si}. Note that the charging term $E_\mathrm{C}\hat{n}^{2}$ is included in $H_\mathrm{junction}$ rather than in $H_\mathrm{env}$, so that the junction Hamiltonian alone makes the $E_\mathrm{J}/E_\mathrm{C}$ competition manifest, while $H_\mathrm{env}$ contains only the dissipative bath.

The voltage drop across the junction is $V_\mathrm{J}=(\hbar/2e)\dot{\hat{\varphi}}$, and the dynamics derived from Eqs.~\eqref{eq:Htotal}--\eqref{eq:Henv} reproduces the resistively and capacitively shunted junction (RCSJ) circuit, with the dissipationless Cooper pair tunneling, the displacement current through $C_\mathrm{J}$, and the dissipative current through $R$ adding up to the bias current.

The derivation rests on four assumptions, which we discuss in the context of our experiment below: (i)~the voltage is well below the gap so that quasiparticle tunneling can be neglected; (ii)~the dimensionless environmental impedance $\rho = R/R_\mathrm{Q}$, with $R_\mathrm{Q}=h/4e^{2}$, is small, $\rho \ll 1$; and (iii)~the junction is overdamped, $\omega_\mathrm{J}/\omega_R \ll 1$, with 
\begin{equation}
    \omega_R = 1/RC_\mathrm{J}\text{ and }\omega_\mathrm{J} = (2e/\hbar)RI_\mathrm{C}.
\end{equation}
(iv)~The current-phase relation is taken to be sinusoidal.

\subsection{Stewart-McCumber parameter}

A useful dimensionless measure of the damping of the junction is the Stewart-McCumber parameter
\begin{equation}
\beta_\mathrm{C} = \frac{2e}{\hbar}\,I_\mathrm{C}\,R^{2}\,C_\mathrm{J}
= \frac{\omega_\mathrm{J}}{\omega_R},
\label{eq:betaC_def}
\end{equation}
with values $\beta_\mathrm{C}<1$ corresponding to the overdamped regime. Using $E_\mathrm{J}=(\hbar/2e)I_\mathrm{C}$, $E_\mathrm{C}=2e^{2}/C_\mathrm{J}$, and $\rho = R/R_\mathrm{Q} = 4Re^{2}/h$, the prefactors $C_\mathrm{J}$ and $I_\mathrm{C}$ in Eq.~\eqref{eq:betaC_def} can be expressed through the energy scales of $H_\mathrm{junction}$, yielding
\begin{equation}
\beta_\mathrm{C} = 2\pi^2\rho^2\,\frac{E_\mathrm{J}}{E_\mathrm{C}}\,.
\label{eq:betaC_EJEC}
\end{equation}
Hence, $\beta_\mathrm{C}$ is determined by the same ratio $E_\mathrm{J}/E_\mathrm{C}$ that controls the charge-phase competition in $H_\mathrm{junction}$, scaled by the squared environmental impedance. In our experiment, $\rho$ is fixed and small, so that the overdamped condition $\beta_\mathrm{C}<1$ is maintained even when $E_\mathrm{J}/E_\mathrm{C}$ becomes substantially larger than unity, consistent with assumption~(iii) of the derivation. This is the curve plotted alongside $E_\mathrm{J}/E_\mathrm{C}$ in Fig.~4(b) of the main text, where both quantities cross unity in the same conductance window near $G_\mathrm{N}\approx G_0$.

\subsection{Cooper pair current as an expansion in $E_\text{J}$}

We treat the Josephson term as the part of the Hamiltonian that couples to $\hat{\varphi}$ in a non-quadratic way, and use it to organize a series expansion of the Cooper pair current. The thermal-equilibrium current is expressed in the interaction picture as
\begin{equation}
I = I_\mathrm{C}\,\Big\langle\,\mathcal{U}^{\dagger}(\infty,t_0)\,\sin[\hat{\varphi}(t_0)]\,\mathcal{U}(\infty,t_0)\,\Big\rangle_{\!\beta},
\label{eq:I_ansatz}
\end{equation}
where the time evolution induced by the Josephson coupling is given by
\begin{equation}
\mathcal{U}(t,t_0) = \mathcal{T}\exp\!\left[\frac{i}{\hbar}\!\int_{t_0}^{t}\!\!\mathrm{d}t'\,E_\mathrm{J}\cos[\hat{\varphi}(t')]\right]\!,
\label{eq:U_evolution}
\end{equation}
with $\mathcal{T}$ the time-ordering operator. The phase $\hat{\varphi}(t)$ evolves under the unperturbed Hamiltonian $E_\mathrm{C}\hat{n}^{2}+H_\mathrm{env}$, which is quadratic in $\hat{n}$, $\hat{\varphi}$, and the bath coordinates. The thermal average $\langle\cdot\rangle_\beta$ is therefore Gaussian, and is taken over the joint equilibrium state of the junction capacitance and the environment. Here and in the following, $\beta=1/k_\text{B}T$ is the inverse temperature. In the limit $t\to\infty$, the initial time $t_0$ becomes irrelevant and is set to zero.

Expanding the time-ordered exponentials in Eq.~\eqref{eq:U_evolution} order by order in $E_\mathrm{J}$, decomposing each $\cos\hat{\varphi}$ and the outer $\sin\hat{\varphi}$ into exponentials $\mathrm{e}^{\pm i\hat{\varphi}}$, and evaluating the resulting Gaussian thermal averages, only odd orders in $E_\mathrm{J}$ contribute and one obtains~\cite{grabert_phase_1998_si}
\begin{align}
I = i\,\frac{I_\mathrm{C}}{2}\sum_{M=1}^{\infty}\!\Big(\frac{i}{2\hbar}E_\mathrm{J}\Big)^{\!2M-1}
&\sum_{\{\zeta,\eta\}}\!\Big(\!\!\prod_{k=1}^{2M-1}\!\eta_k\Big)\zeta_0\nonumber\\[-2pt]
\times\!\int_0^\infty\!\!\!\mathrm{d}t_1\!\dotsi\!\int_0^{t_{2M-2}}\!\!\!&\mathrm{d}t_{2M-1}\,
\exp\!\bigg[\,i\,\frac{2e}{\hbar}V\!\sum_{k=0}^{2M-1}\!\zeta_k t_k\nonumber\\[-2pt]
&\;\;-\sum_{k=1}^{2M-1}\sum_{l=0}^{k-1}\zeta_k\zeta_l\,J[\eta_k(t_k-t_l)]\bigg]\!.
\label{eq:expansion_zeta_eta}
\end{align}
Here, the indices $\zeta_k=\pm 1$ ($k=0,\dots,2M-1$), with the charge-neutrality constraint $\sum_k\zeta_k=0$, originate from the decomposition of the trigonometric functions into exponentials, while $\eta_k=\pm 1$ keeps track of the operator ordering between $\mathcal{U}^\dagger$ and $\mathcal{U}$ in Eq.~\eqref{eq:I_ansatz}. The phase correlation function~\cite{ingold_charge_1992_si}
\begin{equation}
J(t) = 2\!\int_{-\infty}^{+\infty}\!\!\frac{\mathrm{d}\omega}{\omega}\,\frac{\mathrm{Re}\,Z_t(\omega)}{R_\mathrm{Q}}\,
\frac{\mathrm{e}^{-i\omega t}-1}{1-\mathrm{e}^{-\beta\hbar\omega}}
\label{eq:Jt_general}
\end{equation}
encodes the entire effect of the environment (including the contribution from the junction capacitance) on the Cooper pair tunneling, with $Z_t(\omega)$ the impedance seen from the junction. Equation~\eqref{eq:expansion_zeta_eta} sums all orders in $E_\mathrm{J}$ within the model defined by Eqs.~\eqref{eq:Htotal}--\eqref{eq:Henv}. Although the series is generated by an order-by-order expansion of the time-evolution operator, the steps that follow resum it in closed form; the result is therefore non-perturbative in $E_\mathrm{J}$ and is not restricted to the limits $E_\mathrm{J}\ll E_\mathrm{C}$ or $E_\mathrm{J}\gg E_\mathrm{C}$.

\subsection{Phase correlation function in the overdamped regime}

For the $RC$-parallel environment seen from the junction,
\begin{equation}
\frac{\mathrm{Re}\,Z_t(\omega)}{R_\mathrm{Q}} = \frac{\rho}{1+(\omega/\omega_R)^2},
\end{equation}
direct evaluation of Eq.~\eqref{eq:Jt_general} yields a sum of Matsubara contributions plus an exponential and a logarithmic piece~\cite{grabert_phase_1998_si}
\begin{align}
    J(t) &= -2\rho \bigg[ \frac{\pi}{\hbar\beta} |t| + S + \frac{\pi}{2} e^{-\omega_\mathrm{R}|t|} \cot\!\left(\frac{\beta\hbar\omega_\mathrm{R}}{2}\right) \nonumber\\
    &  - \sum_{n=1}^{\infty} \frac{e^{-\nu_n |t|}}{n\left[1 - (\nu_n/\omega_\mathrm{R})^2\right]} + i\frac{\pi}{2}\left(1 - e^{-\omega_\mathrm{R}|t|}\right)\mathrm{sgn}(t) \bigg]
\end{align}
with
\begin{equation}
S = \gamma_\text{E} + \frac{\pi^2\rho}{\beta E_\mathrm{C}} + \psi\!\left(\frac{\beta E_\mathrm{C}}{2\pi^2\rho}\right)\!,
\label{eq:S}
\end{equation}
where $\gamma_\text{E}$ is Euler's constant, $\psi(x)$ is the digamma function, and $\nu_n=2\pi n/\hbar\beta$ are the Matsubara frequencies. The factor~$S$ encodes the temperature-, charging-energy-, and impedance-dependence of the equilibrium phase fluctuations. In our experiment, $\beta$, $E_\mathrm{C}$, and $\rho$ are held fixed across the conductance series, so that $S$ is a constant of the model.

Assuming that the junction is overdamped, i.e.\ $\omega_R\gg\omega_\mathrm{J}$, the terms involving $\mathrm{e}^{-\omega_R|t|}$ can be discarded and we find
\begin{equation}
    J(t) = -2\rho \left[ \frac{\pi}{\hbar\beta} |t| + S - \sum_{n=1}^{\infty} \frac{e^{-\nu_n |t|}}{n\left[1 - (\nu_n/\omega_\mathrm{R})^2\right]} + i\frac{\pi}{2}\,\mathrm{sgn}(t) \right].
    \label{eq:J14}
\end{equation}

Furthermore assuming $\rho\beta E_\mathrm{J}\beta_\text{C}\ll 1$, the Matsubara sum can be discarded as well, and $J(t)$ simplifies to
\begin{equation}
J(t) = -2\rho\!\left[\frac{\pi}{\hbar\beta}|t| + S + i\,\frac{\pi}{2}\,\mathrm{sgn}(t)\right]\!,
\label{eq:Jt_simplified}
\end{equation}

This simplified phase correlation function makes the time integrals analytically tractable with an analytical expression for the Josephson current.

At this point, it is instructive to place these simplifications in the context of the mechanical analog of Josephson junction physics. In this analogy, the phase across the junction takes the role of the position of a dissipative pendulum and the charge takes the role of the conjugate momentum. The overdamped regime then implies $\beta_\mathrm{C} = \omega_\mathrm{J}/\omega_\mathrm{R} < 1$, where $\omega_\mathrm{R}$ plays, in the mechanical analog, the role of the friction constant. From a dynamical point of view, this overdamped regime is characterized by a separation of time scales, where the time scale for the relaxation of position $\omega_\mathrm{R}/\omega_\mathrm{J}^2$ by far exceeds the time scale for the relaxation of momentum $1/\omega_\mathrm{R}$. For this domain, a general theoretical framework has been developed in~\cite{ankerhold_strong_2001_si, ankerhold_overdamped_2004_si} in the form of the so-called quantum Smoluchowski theory. The corresponding evolution equation captures the dynamics of the marginal distribution of position in a time-local diffusion equation with a diffusion coefficient that reduces to the classical, position-independent expression at high temperatures (classical Smoluchowski equation) and to the position-dependent quantum diffusion at low temperatures (quantum Smoluchowski equation). In the low-temperature range, time-locality (Markovianity) is only guaranteed if the quantum time scale $\hbar\beta$ remains smaller than the relevant time scale for the relaxation of position $\omega_\mathrm{R}/\omega_\mathrm{J}^2$. Expressing these conditions for the Josephson junction, one has
\begin{align}
\beta_\mathrm{C} < 1 &\quad\Rightarrow\quad 2\pi^2 \rho^2 E_\mathrm{J} < E_\mathrm{C}, \label{eq:xy}\\
\hbar\beta \ll \omega_\mathrm{R}/\omega_\mathrm{J}^2 &\quad\Rightarrow\quad \beta E_\mathrm{J}\, 2\pi\rho\, \beta_\mathrm{C} < 1. \label{eq:xyz}
\end{align}
Accordingly, the simplifications of $J(t)$ are based on these constraints and thus fall within the domain of the quantum Smoluchowski dynamics. The classical Smoluchowski dynamics leads, for the supercurrent through the Josephson junction, to the Ivanchenko-Zil'berman expression, and its generalization to the quantum regime to the quantum version given in Eq.~\eqref{eq:qIZ} below.

\subsection{Continued fraction and closed form}

Inserting Eq.~\eqref{eq:Jt_simplified} into Eq.~\eqref{eq:expansion_zeta_eta} renders all time integrals analytically tractable. After the integrations, the residual sum over the configurations $\{\zeta_k,\eta_k\}$ admits a one-dimensional Coulomb-gas representation~\cite{grabert_phase_1998_si}, which can in turn be recast as a continued fraction:
\begin{equation}
I = I_\mathrm{C}\,\mathrm{Re}\!\left[\frac{\sin(\pi\rho)}{2\pi\rho}\,
\frac{\mathrm{e}^{-2\rho S}}{\nu+i/\beta E_\mathrm{J}}\,
\cfrac{1}{1+\cfrac{b_1}{1+\cfrac{b_2}{1+\cdots}}}\right]\!,
\label{eq:cfrac}
\end{equation}
with coefficients
\begin{equation}
b_n = \!\left(\frac{\beta E_\mathrm{J}}{2\pi\rho}\right)^{\!2}\!\!
\frac{\sin(\pi\rho n)\sin(\pi\rho(n{+}1))\,\mathrm{e}^{-2\rho S}}{n(n{+}1)(n-i\nu\beta E_\mathrm{J})(n{+}1-i\nu\beta E_\mathrm{J})},
\label{eq:bn}
\end{equation}
and $\nu = V/RI_\mathrm{C}$ the dimensionless voltage. Equation~\eqref{eq:cfrac} is the resummation of the full series in $E_\mathrm{J}$ from Eq.~\eqref{eq:expansion_zeta_eta}, and is therefore valid throughout the crossover rather than only in the limits of small or large $E_\mathrm{J}/E_\mathrm{C}$ where lowest-order treatments apply. It also provides a direct numerical route to evaluating the current: for $\rho\ll 1$, the continued fraction converges rapidly, and only coefficients $b_n$ up to moderately large $n$ are needed.

For $\rho\ll 1$, the sine functions in Eq.~\eqref{eq:bn} can be linearized,
\begin{equation}
\sin(\pi\rho n)\sin(\pi\rho(n+1))\,\mathrm{e}^{-2\rho S}\;\to\;(\pi\rho)^{2}n(n+1)\,\mathrm{e}^{-2\rho S}, \label{eq:sinterm}
\end{equation}
which absorbs the leading effect of the environment into an \emph{effective Josephson energy}
\begin{equation}
E_\mathrm{J}^{*} = E_\mathrm{J}\,\mathrm{e}^{-\rho S}.
\label{eq:EJstar}
\end{equation}
The coefficients in Eq.~\eqref{eq:bn} then reduce to
\begin{equation}
b_n = \!\left(\frac{\beta E_\mathrm{J}^{*}}{2}\right)^{\!2}\!\!\frac{1}{(n-i\nu\beta E_\mathrm{J})(n+1-i\nu\beta E_\mathrm{J})},
\end{equation}
and the continued fraction in Eq.~\eqref{eq:cfrac} can be summed in closed form using the recursion relations of the modified Bessel functions, yielding the central result of the qIZ model:
\begin{equation}
I = \frac{2e}{\hbar}\,E_\mathrm{J}^{*}\,\mathrm{Im}\!\left[\frac{I_{1-i\beta eV/\pi\rho}\!\left(\beta E_\mathrm{J}^{*}\right)}{I_{-i\beta eV/\pi\rho}\!\left(\beta E_\mathrm{J}^{*}\right)}\right]\!,
\label{eq:qIZ}
\end{equation}
where $I_\nu(x)$ is the modified Bessel function of the first kind. The Cooper pair current spectra shown in Fig.~4(a) of the main text are obtained directly from Eq.~\eqref{eq:qIZ}. The switching current $I_\mathrm{S}$ plotted as the purple line in Fig.~2 of the main text is the corresponding maximum of $I(V)$.

Finally, we discuss the range of applicability of the theoretical model. As seen in Fig.~4(b), the condition for overdamped dynamics is always fulfilled within the range of conductances covered in the experiment. The second condition, i.e.\ Eq.~\eqref{eq:xyz}, applies only to conductances $G_\mathrm{N} < 15\,G_0$, so that the spectra up to $G_\mathrm{N} = 28.5\,G_0$ lie formally outside this range. However, in that range we are already well in the domain where the supercurrent grows linearly with $G_\mathrm{N}$, i.e.\ in the phase-dominated but still overdamped regime. While there the Markovianity condition no longer holds in the strict sense, one can assume that in the steady-state regime (long times) the retarded feedback of the electromagnetic reservoir onto the phase dynamics remains small, so that the $I(V)$-curve is well captured by the classical Ivanchenko-Zil'berman expression, in agreement with the experiment (see Fig.~2 and Fig.~4(e)). Since even in this domain $\beta E_\mathrm{J} \gg 1$ as well as $E_\mathrm{J}/\hbar\omega_\mathrm{J} \gg 1$, phase slips via thermal activation or via macroscopic quantum tunneling are negligible, so that quantum effects may only appear via small renormalizations of the Josephson energy (cf.\ Eq.\ \eqref{eq:EJstar}). Nevertheless, a theoretical framework covering not only the crossover from the DCB to the phase-current regime, but also this weak quantum regime is still urgently needed.

\subsection{Limiting cases}

In our experiment, the temperature, charging energy, and environmental impedance are held fixed across the conductance series, so $\beta$, $E_\mathrm{C}$, and $\rho$ — and hence $S$ — are constants. The only parameter that is swept is the Josephson energy through the relation $E_\mathrm{J}\propto G_\mathrm{N}$, which is equivalent to sweeping the dimensionless ratio $E_\mathrm{J}/E_\mathrm{C}$. The two limits of Eq.~\eqref{eq:qIZ} relevant for our experiment are therefore distinguished by the magnitude of $\beta E_\mathrm{J}^{*}\propto G_\mathrm{N}$:

\paragraph{Coulomb blockade limit (DCB regime, small $E_\mathrm{J}/E_\mathrm{C}$).} For $\beta E_\mathrm{J}^{*}\ll 1$, an expansion of Eq.~\eqref{eq:qIZ} to lowest non-trivial order in $E_\mathrm{J}$ yields a Lorentzian peak around zero bias,
\begin{equation}
I = f\,\frac{2\pi^2\rho\,\beta eV}{(\beta eV)^2 + \pi^2\rho^2},
\label{eq:DCB_lorentzian}
\end{equation}
with prefactor
\begin{equation}
f = \frac{\pi e}{\hbar}\,\frac{E_\mathrm{J}^{2}}{E_\mathrm{C}}\,\rho^{2\rho}\!\left(\frac{\beta E_\mathrm{C}}{2\pi^2}\right)^{1-2\rho}\!\mathrm{e}^{-2\rho\gamma_\text{E}}.
\label{eq:f_DCB}
\end{equation}
The quadratic dependence $I\propto E_\mathrm{J}^{2}\propto G_\mathrm{N}^{2}$ is the hallmark of incoherent Cooper pair tunneling in the DCB regime, in agreement with the leading-order $P(E)$-theory result obtained from a perturbative treatment in $E_\mathrm{J}$~\cite{ingold_cooperpair_1994_si}, and matches the observed evolution of $I_\mathrm{S}$ at low conductance in Fig.~2 of the main text.

In the standard $P(E)$-theory~\cite{ingold_charge_1992_si,ingold_cooperpair_1994_si}, the Cooper pair current to lowest order in $E_\mathrm{J}$ reads
\begin{equation}
I_{P(E)}(V) = \frac{\pi e}{\hbar}\,E_\mathrm{J}^{2}\,\bigl[\,P(2eV) - P(-2eV)\,\bigr],
\label{eq:I_PE}
\end{equation}
where $P(E)$ is the probability density for the junction to exchange the energy~$E$ with its electromagnetic environment. It is given by the Fourier transform of the phase correlation function~$J(t)$ defined in Eq.~\eqref{eq:Jt_general}~\cite{ingold_charge_1992_si},
\begin{equation}
P(E) = \frac{1}{2\pi\hbar}\!\int_{-\infty}^{+\infty}\!\!\mathrm{d}t\,
\exp\!\bigl[\,J(t) + iEt/\hbar\,\bigr],
\label{eq:PE_FT}
\end{equation}
so that the same correlation function~$J(t)$ underlies both the qIZ derivation [via Eqs.~\eqref{eq:expansion_zeta_eta} and~\eqref{eq:Jt_simplified}] and the $P(E)$-theory result in Eq.~\eqref{eq:I_PE}. For the ohmic environment with the impedance seen from the junction given at $\omega_R\to\infty$, evaluation of Eq.~\eqref{eq:PE_FT} yields the closed form~\cite{ingold_charge_1992_si,grabert_phase_1998_si}
\begin{equation}
P(E) = \frac{1}{\hbar\Gamma(2\rho)}\!\left(\frac{\pi}{\beta E_\mathrm{C}^{*}}\right)^{\!2\rho-1}\!
\frac{\bigl|\Gamma(\rho+i\beta E/2\pi)\bigr|^{2}}{2\pi}\,
\mathrm{e}^{\beta E/2},
\label{eq:PE_ohmic}
\end{equation}
with $\Gamma(x)$ the gamma function and $E_\mathrm{C}^{*}=E_\mathrm{C}\mathrm{e}^{-2\gamma_\text{E}}/\pi^{2}\rho$ a renormalized charging energy that absorbs the ultraviolet cutoff. Inserting Eq.~\eqref{eq:PE_ohmic} into Eq.~\eqref{eq:I_PE} yields the lowest-order DCB current,
\begin{equation}
I_{P(E)} = f\,\frac{\bigl|\Gamma(\rho-i\beta eV/\pi)\bigr|^{2}}{\Gamma(2\rho)}\,\sinh(\beta eV),
\label{eq:I_DCB_ohmic}
\end{equation}
where the prefactor~$f$ is the same as in Eq.~\eqref{eq:f_DCB}. The qIZ Lorentzian in Eq.~\eqref{eq:DCB_lorentzian} and the $P(E)$-theory result in Eq.~\eqref{eq:I_DCB_ohmic} therefore share the same $E_\mathrm{J}^{2}/E_\mathrm{C}$ scaling, the same temperature exponent $(\beta E_\mathrm{C})^{1-2\rho}$, and the same impedance dependence~\cite{grabert_phase_1998_si}, differing only in the line shape: the qIZ form is a pure Lorentzian, whereas the $P(E)$-theory form carries the gamma-function structure of the full $P(E)$. The two expressions agree precisely close to zero bias voltage and around the peak position to order $\rho^{2}$. This convergence between the two expressions is shown in Fig.~4(e) of the main text, where the qIZ model and the $P(E)$-theory curve with an ohmic environment overlap at low conductance.

\paragraph{Classical limit (DCT regime, large $E_\mathrm{J}/E_\mathrm{C}$).} For $\beta E_\mathrm{J}^{*}\gg 1$, asymptotic expansion of the modified Bessel functions in Eq.~\eqref{eq:qIZ} reduces it to the form of the classical Ivanchenko-Zil'berman result~\cite{ivanchenko_josephson_1969_si} for an overdamped Josephson junction with thermal phase diffusion~\cite{grabert_phase_1998_si}, with $E_\mathrm{J}$ replaced by $E_\mathrm{J}^{*}$. The switching current then scales linearly with $E_\mathrm{J}^{*}\propto G_\mathrm{N}$, as observed at high conductance in Fig.~2 of the main text. The renormalization $E_\mathrm{J}\!\to\! E_\mathrm{J}^{*}$ remains finite at our fixed values of $\beta$, $E_\mathrm{C}$, and $\rho$, so that quantum phase fluctuations persist even in this limit~\cite{ingold_effect_1999_si} and the experimentally extracted critical current does not exactly coincide with the bare Ambegaokar-Baratoff value.

\subsection{Application to the experiment}

In the comparison with the data in the main text, Eq.~\eqref{eq:qIZ} is evaluated with the parameters quoted there: $\Delta_\mathrm{tip}=0.7\,$meV, $\Delta_\mathrm{sample}=0.75\,$meV, $C_\mathrm{J}=1\,$fF, $R_\mathrm{env}=67.32\, \upOmega$, $R_\mathrm{L}=484.1\,\upOmega$, $T=10\,$mK, and a noise broadening of $4.4\,\upmu$V. The only parameter varied across the conductance series is the Josephson energy through the relation $E_\mathrm{J}\propto G_\mathrm{N}$; all environmental parameters are held fixed, consistent with the experimental situation in which only the tip-sample distance is changed. The transition $E_\mathrm{J}/E_\mathrm{C}\approx 1$ is then found to occur near $G_\mathrm{N}\approx G_0$, in agreement with the experimentally observed crossover from the quadratic to the linear evolution of $I_\mathrm{S}$ in Fig.~2 of the main text. The corresponding evolution of the Stewart-McCumber parameter, Eq.~\eqref{eq:betaC_EJEC}, keeps $\beta_\mathrm{C}<1$ across the entire range despite $E_\mathrm{J}/E_\mathrm{C}$ exceeding unity at high conductance, as plotted in Fig.~4(b) of the main text. This is consistent with assumption~(iii) of the derivation.

Two assumptions of the qIZ model deserve a comment in the context of our experiment. First, the environment is reduced to a purely ohmic resistance, so that the resonances of the realistic scan-head impedance are not captured \cite{zeng_enhancing_2026_si}. This affects the spectral shape at higher voltages but not the overall evolution of $I_\mathrm{S}$. Second, the current-phase relation is assumed sinusoidal, which becomes a less accurate approximation in the high-conductance regime where multiple high-transmission channels contribute (see Fig.~3 of the main text). Within these limitations, the qIZ model provides a quantitatively consistent description of the full crossover from the DCB to the DCT regime, as discussed in Fig.~4 of the main text.

In the qIZ model, a single resistance both sets the dc voltage division, $V_\mathrm{J}=V_\mathrm{B}-R\,I$, and constitutes the dissipative environment $Z(\omega)=R$ to which the phase couples~\cite{grabert_phase_1998_si, ingold_charge_1992_si}. In the present experiment these two roles are played by different, independently determined elements. The line resistance $R_\mathrm{L}=484.1\,\upOmega$ is a dc quantity fixed by the wiring and filtering~\cite{assig_10_2013_si,zeng_enhancing_2026_si}, which is used only to convert the applied bias to the junction voltage. The environment governing the phase dynamics is the impedance seen by the junction at its own dynamical frequencies $\omega_R=1/R_\mathrm{env}C_\mathrm{J}$ and $E_\mathrm{C}/\hbar$, approximated by an effective ohmic resistance $R_\mathrm{env}=67.32\,\upOmega$. At these frequencies the junction and stray capacitances shunt the distant line resistance, so that $R_\mathrm{env}$ is set by the local environment and is generally distinct from $R_\mathrm{L}$. The theoretical qIZ circuit is recovered for $R_\mathrm{L}=R_\mathrm{env}$.

\section{Phase Rotation per Tunneling Particle}

The tunneling current and the bias voltage are conjugate measurements of the rate of charge transfer and the rate of phase rotation across the junction, respectively~\cite{tinkham_introduction_1996_si}. The current measures the change of the particle number per unit time,
\begin{equation}
I=2e\,\frac{d n}{dt},
\label{eq:I_def}
\end{equation}
where the factor of $2e$ accounts for the charge of each Cooper pair. The phase evolves according to the second Josephson relation~\cite{josephson_possible_1962_si,tinkham_introduction_1996_si},
\begin{equation}
\frac{d\phi}{dt}=\frac{2eV}{\hbar}.
\label{eq:phi_def}
\end{equation}
Combining the two relations gives the phase rotation per tunneling Cooper pair:
\begin{equation}
\frac{\Delta\phi}{\Delta n}=\frac{d\phi/dt}{dn/dt}=\frac{(2eV/\hbar)}{I/(2e)}=\frac{4e^2 V}{\hbar I}.
\label{eq:dphi_dn_general}
\end{equation}
Evaluating Eq.~\eqref{eq:dphi_dn_general} at zero bias and using $I/V=G_\text{ZB}$, together with the conductance quantum $G_0=2e^2/h$, we obtain
\begin{equation}
\frac{\Delta\phi}{\Delta n}=\frac{4e^2}{\hbar G_\text{ZB}}=4\pi\,\frac{G_0}{G_\text{ZB}},
\label{eq:dphi_dn}
\end{equation}
which is the relation given in the main text. Equation~\eqref{eq:dphi_dn} is a simple consequence of the Josephson relations combined with the definition of the conductance: it relates how much phase rotates per tunneling event to the inverse of the zero-bias conductance, expressed in units of $G_0$.

In the DCB regime, the zero-bias conductance is small, $G_\text{ZB}\ll G_0$, so $\Delta\phi/\Delta n\gg 1$: the phase rotates many times between tunneling events, consistent with the picture of incoherent Cooper pair tunneling driven by phase fluctuations. In the DCT regime, the zero-bias conductance is large, $G_\text{ZB}\gg G_0$, so $\Delta\phi/\Delta n\ll 1$: the phase barely changes per tunneling event, consistent with phase-coherent transport. The transition between these two limits, shown in Fig.~4(c) and (d) of the main text, tracks the emergence of phase coherence as $E_\text{J}/E_\text{C}$ is increased. The horizontal dashed line in Fig.~4(c) corresponds to the experimental sensitivity limit set by the smallest measurable $\Delta\phi/\Delta n$ at the largest $G_\text{ZB}$ accessible in the mK-STM.

\end{document}